\documentclass[graybox]{svmult}

\usepackage{type1cm}        
\usepackage{makeidx}         
\usepackage{graphicx}        
\usepackage{multicol}        
\usepackage[bottom]{footmisc}

\usepackage{newtxtext}       %
\usepackage[varvw]{newtxmath}       

\makeindex             

\begin{document}

\title*{An Unfinished Collaboration \\
with   A. A. Starobinsky}
\author{Thibault Damour\orcidID{0000-0003-4519-6265}}
\institute{Institut des Hautes Etudes Scientifiques, 35 route de Chartres, 91440 Bures sur Yvette, France \email{damour@ihes.fr}}
%
%
\maketitle

\abstract*{The present text summarizes some of the results obtained in 2008 during the initial stages of a collaboration
with Starobinsky which remained unfinished. The collaboration was an attempt to apply the stochastic approach to
infrared (IR) quantum-gravity effects in an inflationary spacetime.}

\abstract{The present text summarizes some of the results obtained in 2008 during the initial stages of a collaboration
with Starobinsky which remained unfinished. The collaboration was an attempt to apply the stochastic approach to
infrared (IR) quantum-gravity effects in an inflationary spacetime.}

\section{Introduction}
\label{sec:1}
My dear friend Alexei Alexander Starobinsky (henceforth Alyosha) visited the Institut des Hautes Etudes Scientifiques (IHES) 
in October 2008. During this visit, we started a collaboration on the impact of infrared quantum gravity effects on the global  structure of
de Sitter (or inflationary) spacetimes. I had always hoped that Alyosha would come back at IHES to allow us to continue our discussions, and to
hopefully finalize the preliminary results we had reached in October 2008. Alas this did not happen. Though I had the pleasure
of meeting again Alyosha in person during some later visits of mine in Moscow (notably during the 2019 conference dedicated to 
the 100th anniversary of Khalatnikov), and though Alyosha kindly agreed to speak (remotely) during the 2021 ``Damour Fest"
at IHES, we were both dragged away from our collaborative project by other scientific pursuits, and we never finalized our collaboration.
As an homage to him, and as a token of my admiration for all the deep and beautiful ideas he contributed to physics, the present
text will present the gist of our discussions,  as they were left at the time (together with some additional considerations
on Brownian motion in the space of 3-by-3 matrices). Hopefully, the approach we pursued (crucially based on
Alyosha's stochastic approach to inflation) has still some value, in spite of 
the many developments that have since happened (and are still happening) in the study of infrared effects in de Sitter
spacetimes.  Before entering into details, let me mention that the stochastic approach to inflation is the idea, pioneered in 
Ref. \cite{Starobinsky:1982ee} (see also \cite{Vilenkin:1983xq,Linde:1986fd}, as well as \cite{Starobinsky:1986fx,Starobinsky:1994bd}),
that the infrared part of a quantum fluctuating field during inflation may be considered as a classical (i.e., $c$-number) space-dependent 
stochastic field satisfying a local (i.e., without spatial derivatives) evolution equation where the classical evolution is completed by an
additional Langevin-like fluctuating term (coming from the higher-frequency quantum fluctuations). For  recent reviews
of the stochastic approach to inflation see Ref. \cite{Cruces:2022imf,Woodard:2025cez}.  
For  entries into the still growing literature on infrared
effects in de Sitter (or inflationary) spacetimes, see \cite{Polyakov:2012uc,Anderson:2017hts,Akhmedov:2019cfd,Akhmedov:2019esv,Cohen:2020php,Miao:2024shs,Woodard:2025smz}.
The last two references, of which I became aware while writing the present account, have close links with the aim of my unfinished 
collaboration with Alyosha. I leave to the future to decide whether they make it irrelevant by superseding it.

\section{First starting point}
\label{sec:2}
The first starting point of our discussions was the following  classic result of Alyosha.
In 1983, Alyosha derived the asymptotic structure, at large times, of the general solution
of Einstein's equations in presence of a (positive) cosmological constant $\Lambda= 3 H^2$ \cite{Starobinsky:1982mr}
(in $3+1$ dimensions, and when neglecting other types of stress-energy tensor), namely (in synchronous gauge)
\begin{equation} \label{ds2}
ds^2= - dt^2 + e^{2 H t} \hat g_{ij}({\bf x}, t) dx^i dx^j\,,
\end{equation}
where the rescaled spatial metric $\hat g_{ij}({\bf x}, t)$ admits an expansion in powers of $ e^{- H t} $ of the form
\begin{equation} \label{exp}
\hat g_{ij}({\bf x}, t) = a_{ij}({\bf x}) +  b_{ij}({\bf x})  e^{-2 H t} +  c_{ij}({\bf x})  e^{-3 H t}+ \cdots \,.
\end{equation}
One can further impose the unimodularity condition $ \det  a_{ij}({\bf x})=1$. The latter expanded solution\footnote{In an anti-de-Sitter
context, it would be called a Fefferman-Graham expansion.} is general in that it
contains four physical arbitrary functions of three variables:  there are two arbitrary functions in the unimodular spatial metric 
 $a_{ij}({\bf x})$ (considered modulo 3-diffeomorphisms); the function  $b_{ij}({\bf x})$ is determined by the
 Ricci tensor of $a_{ij}({\bf x})$
 (namely $H^2 b_{ij}({\bf x})= R_{ij}[a]-\frac14 R[a] a_{ij}$); and the other two arbitrary functions of three variables are
 contained in $ c_{ij}({\bf x})$ (which must satisfy $c_i^i =0$ and $\nabla^j c_{ij}=0$). [The latter constraints use $a_{ij}({\bf x})$
 as background metric.] Terms beyond $ c_{ij}({\bf x}) $ in the expansion \eqref{exp} are fully determined by  $a_{ij}({\bf x})$ and 
 $ c_{ij}({\bf x}) $.
 
 The expansion \eqref{exp} is valid in many spatial coordinate systems. E.g., one recovers the global de Sitter metric
 with $ \hat g_{ij}({\bf x}, t) dx^i dx^j = H^{-2} e^{- 2 H t} \cosh^2 H t d \Omega_3^2$ (where $d \Omega_3^2$ is a round
 3-sphere metric), which leads to $ a_{ij}({\bf x}) dx^i dx^j= \frac14  H^{-2}  d \Omega_3^2$
 and  $b_{ij}({\bf x})= 2  a_{ij}({\bf x})$. Alternatively, the Poincar\'e-patch case
 corresponds to $ a_{ij}({\bf x})= \delta_{ij}$ and $ b_{ij}({\bf x})= 0$.

The crucial point here is that though this general solution exhibits a phenomenon of local isotropization,
dubbed the cosmic no-hair behavior (with an exponentially
vanishing Weyl tensor, $C_{\mu \nu \kappa \lambda} C^{\mu \nu \kappa \lambda} \propto  e^{-6 H t}$), its global spatial
geometry is conformal, at large times, to an arbitrary (unimodular) 3-metric $a_{ij}({\bf x}) = \lim_{t \to \infty} \hat g_{ij}({\bf x}, t)$.
The conformal spatial geometry defined by  $a_{ij}({\bf x})$ is globally inhomogeneous and anisotropic. It can be arbitrarily different
from the round sphere obtained for a global de Sitter spacetime. On the other hand, it is time-independent.

\section{Second starting point}
\label{sec:3}
As just reviewed, in a classical context the large-time limit of the conformal spatial metric exists and defines a globally inhomogeneous
and anisotropic, but time-independent geometry. The second basic starting
point of my collaboration with Alyosha was to extend his stochastic approach to inflation to an approximate 
description of the evolution of $\hat g_{ij}({\bf x}, t) $ in a quantum context, and to argue that a suitably coarse-grained 
version of the quantum $\hat g_{ij}({\bf x}, t) $ would define a stochastically evolving (unimodular) 3-metric $a_{ij}({\bf x},t)$
becoming exponentially anisotropic as $t \to +\infty$.

Let us first recall another classic result of Alyosha, concerning the production of ``relict gravitational radiation" from quantum
gravitational inflationary fluctuations \cite{Starobinsky:1979ty}.  Linearized quantum gravitational  fluctuations in a de Sitter
background, i.e. small fluctuations in the (conformally rescaled) spatial metric of the type,
\begin{equation} \label{gij}
 \hat g_{ij}({\bf x}, t) = \delta_{ij} + h_{ij} ( {\bf x},t)  \,,
\end{equation}
can be Fourier-decomposed as (with $\kappa \equiv \sqrt{32 \pi G}$)
\begin{equation} \label{hij}
 h_{ij} ( {\bf x},t) = \sum_e \kappa \int \frac{d^3k}{(2 \pi)^{\frac32}} e_{ij}  a_{\bf k} e^{ i {\bf k} \cdot {\bf x}} \frac{\chi_{\bf k}(\eta)}{a(\eta)} +{\rm h.c.}
\end{equation}
Here, we assumed for simplicity that we work in a Poincar\'e patch, and we introduced the conformal time $\eta$, 
varying between $- \infty$ and $0^-$, and such that 
\begin{equation}
a= e^{H t} = -\frac1{ H \eta}\,.
\end{equation}
In Eq. \eqref{hij}, $e_{ij}$ labels two different normalized ($e_{ij}e_{ij} =1$) transverse-traceless polarization tensors, 
and the corresponding mode functions\footnote{For simplicity we leave implicit the extra label, say $e= +, \times$, on the mode functions, and on the annihilation operators.} $\chi_{\bf k}(\eta)$ 
satisfy the equation (with $'=d/d\eta$)
\begin{equation}
\chi_{\bf k}''(\eta)+ \left( k^2 - \frac{a''}{a} \right)  \chi_{\bf k}(\eta)=0\,,
\end{equation}
whose solution (corresponding to the Bunch-Davies vacuum) is
\begin{equation}
\chi_{\bf k}(\eta)= \frac{e^{- i k \eta}}{\sqrt{2 k}} \left(1 - \frac{i}{k \eta}  \right) \,.
\end{equation}
Finally, $a_{\bf k}$ is a corresponding quantum annihilation operator satisfying the canonical commutation relation
\begin{equation}
[ a_{\bf k}\, ,  a^{\dagger}_{\bf k'}] = \delta( {\bf k}- {\bf k'})\,,
\end{equation}
and defining the Bunch-Davies vacuum through $ a_{\bf k} | 0\rangle=0$.

A first way to see the physics associated with the quantum gravitational field \eqref{hij} is to compute
the VEV of $h_{ij}^2 ( {\bf x},t)$, namely 
\begin{equation}
\langle 0|h_{ij}^2 ( {\bf x},t)  | 0\rangle=2  \kappa^2 \left(\frac{H}{2 \pi}   \right)^2 \int_{k_{\rm min}}^{k_{\rm max}} \frac{dk}{k} (1+ (k \eta)^2)\,.
\end{equation}
The latter integral is both UV and IR divergent. The UV (power-law) divergence is the usual  local-Minkowski divergence of the two polarization
 modes of the gravitational field (with $ h_{ij}(L) \sim \kappa/L$ over a small (sub-Hubble) spatial length $L \ll H^{-1}$).
 On the other hand, the IR (logarithmic) divergence involves the lower bound $k_{\rm min}= H a(\eta_0)$ corresponding to the Hubble scale of some initial time, $\eta_0$. 
 When integrated over the interval of momenta $k_{\rm min}< k < H a(\eta)$ (where the upper bound 
 $H a(\eta)$ corresponds to the Hubble scale at the considered current time, $\eta$), the contribution of the IR logarithm then gives (using $a(\eta)= e^{H t}$)
 \begin{equation} \label{h2IR}
\langle 0|h_{ij}^2 ( {\bf x},t)  | 0\rangle_{\rm IR} \approx 2 \kappa^2 \left(\frac{H}{2 \pi}   \right)^2 \ln \frac{a(\eta)}{a(\eta_0)}= 2 \kappa^2 \left(\frac{H}{2 \pi}   \right)^2 H \,(t-t_0)\,.
\end{equation}
We see on the latter equation that the IR part of the quantum gravitational fluctuations leads to a linear-in-time wandering
of the squared metric fluctuation. We recognize a Brownian motion of $h_{ij}(t)$ due to quantum modes
coming from the UV, being redshifted by the expansion, and passing through the Hubble scale at time $t$.
Essentially, Eq. \eqref{h2IR} says that each one of the two polarization modes contributes a random amplitude 
$ \pm \kappa \frac{H}{2 \pi} $, during each time interval $\Delta t = H^{-1}$.

\section{Towards a stochastic approach to super-Hubble quasi-classical IR fluctuations of the de Sitter metric}
\label{sec:4}
Alyosha's stochastic approach to quantum fluctuating light scalar fields (with mass $m \ll H$) during inflation consists
in representing the Heisenberg operator of the quantum field $\phi$ as \cite{Starobinsky:1982ee,Starobinsky:1986fx,Starobinsky:1994bd}
\begin{equation} \label{phidecomp} 
\phi({\bf x},t)= \bar \phi({\bf x},t)+ \int \frac{d^3k}{(2 \pi)^{\frac32}} \theta(k-\epsilon a(t) H) \left[  a_{\bf k}   \phi_{\bf k}(t) e^{ i {\bf k} \cdot {\bf x}} +{\rm h.c.} \right]\,.
\end{equation}
Here $\bar \phi({\bf x},t)$ is a coarse-grained version of  $\phi({\bf x},t)$, i.e. averaged over a constant physical volume
larger (given some $\epsilon < 1$) than a Hubble volume. The second term in Eq. \eqref{phidecomp}, which involves
a step function $\theta(x)$ and annihilation/creation operators $a_{\bf k}, a^{\dagger}_{\bf k}$, represents the short-wavelength counterpart of $\phi({\bf x},t)$. The presence of the step
function means that quantum modes of wave number $k$ coming from the UV get redshifted by the expansion, $k^{\rm phys}= \frac{k}{a(t)}$, and thereby continuously transit from the quantum high-frequency part of $\phi({\bf x},t)$ to the low-frequency part $\bar \phi({\bf x},t)$ as $k^{\rm phys}= \frac{k}{a(t)}$ becomes smaller than $\epsilon H$. The basic idea is to approximate the low-frequency part 
$\bar \phi({\bf x},t)$ by a classical stochastic field which can wander to large values in field space, contrary to the high-frequency quantum part, which is treated as staying ``small" (after suitably subtracting UV divergences). When taking the time derivative
of Eq. \eqref{phidecomp}, the derivative of the step function generates a term in $ \partial\phi({\bf x},t)/\partial t$ of the form
$ -  f({\bf x},t)$, where
\begin{equation}
f({\bf x},t) \equiv  \epsilon a(t) H^2\int \frac{d^3k}{(2 \pi)^{\frac32}} \delta(k-\epsilon a(t) H) \left[  a_{\bf k}   \phi_{\bf k}(t) e^{ i {\bf k} \cdot {\bf x}} +{\rm h.c.} \right]\,.
\end{equation}
The term $ + f({\bf x},t)$ then must appear as an additional term in the time derivative of the quasi-classical low-frequency field 
$\bar \phi({\bf x},t)$. The main technical feature of the stochastic approach is then to consider  the term $ + f({\bf x},t)$
as a Langevin-type stochastic force driving (in addition to classical terms) the evolution of $\bar \phi({\bf x},t)$. E.g., in the
case of a scalar field with self-interacting potential $V(\phi)$, one gets the approximate first-order evolution equation
\begin{equation}
\frac{\partial \bar \phi({\bf x},t)}{\partial t} = - \frac1{3H} V'(\bar \phi) + f({\bf x},t) \,,
\end{equation}
where $ \bar \phi({\bf x},t)$ is treated as a classical stochastic variable, and $ f({\bf x},t)$ as a random Gaussian noise,
having the statistical properties defined by the quantum averages of $a_{\bf k}$ and  $a^{\dagger}_{\bf k}$.
Refs. \cite{Starobinsky:1994bd,Woodard:2025cez} have shown that this stochastic approach does correctly describe IR
effects in various field models. See also Ref. \cite{Kiefer:1998qe} for a discussion of why the low-frequency quantum
fluctuations can be treated as classical stochastic variables.

The aim of our unfinished collaboration was to use this stochastic approach to derive an approximate evolution of the 
low-frequency part (say  $ \bar g_{ij}({\bf x}, t)$) of the conformal metric  $\hat g_{ij}({\bf x}, t)$. We  explored several 
possibilities for defining $ \bar g_{ij}({\bf x}, t)$ and for writing down its Langevin-type evolution equation. 
Let me mention one of them.

If we start from the $3+1$ decomposition of Einstein's equations (with $\Lambda$) in a synchronous (or Gauss) gauge, and rewrite them
in terms of  $\hat g_{ij}({\bf x}, t) \equiv a^{-2}  g_{ij}({\bf x}, t) $, with $a(t) = e^{H t}$, they read (with $ \hat K  \equiv  \hat K^s_{\; s}$)
 \begin{eqnarray}
&&  \hat g^{is} \partial_t \hat g_{sj} = 2 {\hat K}^i_{\; j}\,,  \nonumber \\
&& \partial_t  \hat K^i_{\; j} + 3 H \hat K^i_{\; j}+ H \hat K \delta^i_{\; j} + \hat K  \hat K^i_{\; j}+ a^{-2}{\hat R}^i_{\; j}=0\,.
\end{eqnarray}

Let us start, around some spatial point, and at some time, with a local value of the metric of the type
\begin{equation} \label{gdecomp2}
 \hat g_{ij}({\bf x}, t) = \bar g_{ij}({\bf x}, t) + h^{\rm UV}_{ij} ( {\bf x},t) \,,
 \end{equation}
 where the initial value of the low-frequency part is the flat Euclidean metric, $ \bar g_{ij}({\bf x}, t_0)= \delta_{ij}$,
 while  the high-frequency part reads
\begin{equation}
  h^{\rm UV}_{ij} ( {\bf x},t) =  \sum_e \kappa \int \frac{d^3k}{(2 \pi)^{\frac32}}\theta(k-\epsilon a(t) H) \left[ e_{ij}  a_{\bf k} e^{ i {\bf k} \cdot {\bf x}} \frac{\chi_{\bf k}(\eta)}{a(\eta)} +{\rm h.c.} \right]\,.
 \end{equation}
Taking the time derivative of the step function in $ h^{\rm UV}_{ij} ( {\bf x},t) $ suggests that the low-frequency
part will approximately evolve as
\begin{equation} \label{linearLangevin}
\frac{\partial \bar g_{ij}({\bf x}, t)}{\partial t} \approx f_{ij}({\bf x},t)\,,
\end{equation}
where the Langevin force $f_{ij}({\bf x},t)$ (induced by the time derivative of $ h^{\rm UV}_{ij} ( {\bf x},t) $) reads
 \begin{equation}   \label{fij}
f_{ij}({\bf x}, t) \equiv   \epsilon \kappa a(t) H^2 \sum_e \int \frac{d^3k}{(2 \pi)^{\frac32}} \delta(k-\epsilon a(t) H) \left[  
 e_{i j}  a_{\bf k} e^{ i {\bf k} \cdot {\bf x}} \frac{\chi_{\bf k}(\eta)}{a(\eta)} +{\rm h.c.} \right]\,.
 \end{equation}
The trace,  $ \delta^{ij} f_{ij}$, of this Langevin force vanishes, and its two-point correlation has
(when considering for simplicity a fixed spatial point\footnote{Alyosha told me that he expected the (physical) spatial correlation  length 
$R_c$ to be (asymptotically) of (logarithmic) order $\ln (H R_c) \sim (\kappa H)^{-1}$.})
 the following structure:
\begin{equation}
\langle f_{ij}(t) f_{kl}(t')\rangle \propto  (\delta_{ik} \delta_{jl}+ \delta_{il} \delta_{jk} - \frac23 \delta_{ij} \delta_{kl}) \left(\frac{\kappa H}{2 \pi}   \right)^2 H  \delta(t-t')\,.
\end{equation}
When going beyond the approximation of linear stochastic perturbations of $ \bar g_{ij}({\bf x}, t)$  
around an Euclidean metric $\delta_{ij}$, described by Eq. \eqref{linearLangevin}, it seems consistent to define 
the nonlinear stochastic evolution
of $ \bar g_{ij}({\bf x}, t)$ by requiring that $ \bar g_{ij}({\bf x}, t)$ stays  unimodular, and satisfies a  
{\it geometrically-defined} first-order stochastic evolution equation in the space, say $S$, of unimodular, 
positive-definite symmetric 3-by-3 matrices. 

Let us recall that the space of unimodular, positive-definite symmetric 3-by-3 matrices  $ \bar g$  is a 5-dimensional 
Riemannian space, endowed with the metric 
\begin{equation} \label{metricS}
ds^2 = {\rm Tr} [ ( \bar g)^{-1} d  \bar g]^2= \bar g^{ik} \bar g^{jl} d\bar g_{ij} d \bar g_{kl}, 
\end{equation}
and that it is equivalent to the symmetric space $S \equiv SL(3)/SO(3)$.  The equivalence between these two 5-dimensional spaces is
made particularly clear by parametrizing  the matrix $ \bar g= ( \bar g_{ij}({\bf x}, t)) $ by means of an Iwasawa decomposition $ K A N$ 
of the group $SL(3)$. Namely,  a generic
group element $\gamma \in  SL(3)$ can be uniquely decomposed as the product $k a n$ of three 3-by-3 matrices, 
with: $ k \in SO(3)$;  $a$ being diagonal, positive,  and unimodular;
and with $n$ being upper triangular (with ones on the diagonal). The coset $S \equiv SL(3)/SO(3)$ is described by the $ a n$ part, 
which is in one-to-one correspondence with  $ \bar g= \gamma^t \gamma= n^t a^t a n = n^t A n$ ($t$ denoting a transpose,
and $A=a^t a$ being diagonal, positive and unimodular).
More explicitly, we can write at each spatial point ${\bf x}$
\begin{equation} \label{iwasawa}
 \bar g_{ij}( t)= \sum_{a=1}^3 N^a_{\; \: i} e^{ 2 \beta_a} N^a_{\; \; j}\,; {\rm with} \; \beta_1+ \beta_2+\beta_3=0\,.
\end{equation}
Here, $N^a_{\; \: i}$ denotes the components of the upper triangular matrix $n$, while $ A= a^t a = {\rm diag}[e^{2 \beta_a}]$.
The explicit expression of the metric $ds^2$ in the Iwasawa coordinates $\beta_a$, $N^a_{\; \: i}$ can
be found in \cite{Henneaux:1981su}.

The infinitesimal neighbourhood of $ \bar g_{ij}= \delta_{ij}$, say $ \bar g_{ij}= \delta_{ij}+ \epsilon f_{ij}$ (with $f_{ij}$ symmetric and traceless),
is parametrized by $\beta_a \to \epsilon \beta_a$ and $N^a_{\; \; i}= \delta_{ai} + \epsilon n_{ai}$, with $n_{ai}$ strictly
upper triangular (i.e. with $n_{12}, n_{13}, n_{23}$ as only non-zero elements). The correspondence between $ f_{ij}$
and $\beta_a, n_{ai}$ is $f_{aa}= 2 \beta_a$, $f_{12}= n_{12}(= f_{21}), f_{13}=n_{13}, f_{23}=n_{23}$.
The Langevin term \eqref{fij} in the evolution of $ \bar g_{ij}( t)$ near $\delta_{ij}$ then defines a corresponding Langevin
term in the evolution of the coordinates,  $\beta_a$, $n_{12}, n_{13}, n_{23}$, of the coset element $ a n$ near the identity. 

The issue is then to define, at the nonlinear level (when $\bar g$ starts wandering away from the identity),
a definite stochastic evolution of $\bar g$ over the curved manifold $S$. The classic work of It\^o \cite{Ito1950}
has shown how to define stochastic differential equations on a curved manifold in a geometrically covariant way,
i.e. independently of the choice of coordinates on the considered manifold. In the case of the coset space $S$, the
five numbers  $\beta_1$,  $\beta_2$, $n_{12}, n_{13}, n_{23}$ parametrizing $\bar g$ are particular coordinates.
One wants to consider a first-order stochastic differential system that is equivariant under the choice of parametrization of the elements of $S$. Indeed, one might have used coordinates on $S$ based, e.g., on a Gauss decomposition of  the 
matrix $\bar g_{ij}$, as used, e.g., in \cite{Damour:2011yk}.

 It\^o found that a 
geometrically covariant stochastic  differential equation for the motion of the coordinates $X^{\mu}$
of a point  on a Riemannian manifold (with metric tensor $G_{\mu \nu}(X)$ in the coordinates $X^{\mu}$)
must involve an additional (non-covariant) drift vector. The most natural Brownian motion equation for $X^{\mu}(T)$
 (with a suitable scaled time variable $T$) reads
\begin{equation} \label{brownian}
d X^{\mu}(T) = e^{\mu}_a(X)\, dw^a(T) -\frac12 G^{\alpha \beta} \Gamma^{\mu}_{\alpha \beta} dT\,,
\end{equation}
where $ e^{\mu}_a(X)$ is a local orthonormal frame (``rep\`ere mobile", satisfying 
$G_{\mu \nu}  e^{\mu}_a  e^{\nu}_b=\delta_{ab}$), $dw^a(T)$ a normalized white noise
in the tangent space ($ \langle dw^a(T) dw^b(T') \rangle= \delta^{ab}\delta(T-T')$) , 
and $\Gamma^{\mu}_{\alpha \beta}$ the Christoffel coefficients of $G_{\mu \nu}(X)$. 
The diffusion (or Fokker-Planck) equation describing the evolution of the (one-point) probability
distribution function (pdf)  $\rho(X)$ then reads
\begin{equation} \label{heat}
 \frac{ \partial \rho(X)}{\partial T} = \frac12 \Delta_G  \, \rho(X)\,,
  \end{equation}
  where $\Delta_G $ denotes the Laplace-Beltrami operator defined by the  metric $G_{\mu \nu}(x)$.

It was argued above that the IR quantum fluctuations
of the spatial metric $ \bar g_{ij}({\bf x}, t)$, when starting from the flat metric $\delta_{ij}$, satisfy
 $ \frac{\partial \bar g_{ij}(t)}{\partial t} \approx f_{ij}(t) $ with a white noise $f_{ij}$ defined by Eq. \eqref{fij}.
 Viewing  $ \bar g_{ij}({\bf x}, t)$, at each spatial point ${\bf x}$, as a point $X^\mu$ in the 5-dimensional symmetric space
 $S \equiv SL(3)/SO(3)$ (endowed with the metric \eqref{metricS}), it seems
 natural to assume that, at the nonlinear level (when $\bar g$ starts wandering away from the identity),
  the stochastic wandering of  $ \bar g_{ij}({\bf x}, t)$ over the 5-dimensional symmetric space $S$
  will be described by the simple geometric Brownian motion \eqref{brownian}, with associated diffusion equation
   \begin{equation} \label{heateq}
 \frac{ \partial \rho(\bar g)}{\partial T} = \frac12 \Delta_S  \, \rho(\bar g)\,,
  \end{equation}
  for the one-point pdf of $\bar g$ on $S \equiv SL(3)/SO(3)$. Here $\Delta_S $ denotes the Laplace-Beltrami operator
  on $S$, endowed with the metric \eqref{metricS}.
 [We also wondered whether in Eq. \eqref{heateq} $\Delta_S$ should be augmented by a scalar-curvature
term, say $\Delta_S + \xi R_S $ with $\xi =O(1)$.] The heat equation \eqref{heateq} involves a 
suitably rescaled (dimensionless)  time variable, 
\begin{equation}\label{T}
T \sim  \left(\frac{\kappa H}{2 \pi}   \right)^2 H t \,. 
\end{equation}

While writing up the results of my discussions with Alyosha, I became aware of the large mathematical
literature discussing Brownian motions on symmetric spaces. For entries in the literature see notably Refs.\cite{Malliavin1974} 
and \cite{Anker1992}. 

A remarkable fact is that the long-time behavior of a Brownian motion on a non-compact
symmetric space (such as $S$) exhibits special features linked to the negative sectional curvature of such spaces.
Namely, when using an Iwasawa parametrization, say $ \bar g= n^t a^t a n = n^t A n$,
heat diffusion  in  symmetric spaces $G/K$ of rank $\ell \geq 2$ asymptotically (for large times $T$) proceeds in
the following way: (i) on the one hand, the off-diagonal component $n(T)$  (parametrized by $N^a_{\;\; i}$
in Eq. \eqref{iwasawa}), tends to some (random)
finite limit as $T \to +\infty$; while, (ii) on the other hand, the logarithm $\alpha = \ln A$ of the diagonal component
(equal to $\alpha= {\rm diag} (2 \beta^a)$ in the notation of Eq. \eqref{iwasawa})
asymptotically proceeds in a fixed direction (Weyl vector) in the linear Cartan space, and with a 
fixed, finite speed. 

 In our case, $S=G/K$ with $G=SL(3, R)$
and $K=SO(3)$,  the rank (i.e. the number of independent $\beta$'s) is $\ell=2$. Actually, in this case, one
can explicitly write down the solution of the stochastic differential equation satisfied by the five  Iwasawa coordinates
$N^a_{\;\; i}$ and $\beta_a$ (with the  constraint $\beta_1+\beta_2 +\beta_3 =0$), 
in terms of six  Wiener processes $w_I(T)$, $I=1,2,\cdots,6$, as was done in 
the recent Ref. \cite{Matsumoto2022}. [One must subtract from the first three Wiener processes 
in  Section 3.2 of \cite{Matsumoto2022}
their trace $\frac13(w_1+w_2 +w_3)$, i.e. replace $w_a$ by $\bar w_a= w_a-\frac13(w_1+w_2 +w_3)$].
This leads to a result of the following form for the asymptotic (large $T$) stochastic evolution of the metric $ \bar g_{ij}(T)$
(considered at each given spatial point $\bf x$):
 \begin{equation}
 \bar g_{ij}(T) =\sum_{a=1}^3 N^a_{\; \: i}(T) e^{ 2 \beta_a(T)} N^a_{\; \; j}(T)\,,
  \end{equation}
where the three diagonal (logarithmic) components are given by simple explicit expressions involving Wiener processes
(i.e. Brownian motions), namely
 \begin{eqnarray}
  2 \beta_1(T) &=&  2 \beta_1(0)+ \frac12 T + \bar w_1(T), \nonumber \\
   2 \beta_2(T) &=&  2 \beta_2(0)+  \bar w_2(T), \nonumber \\
     2 \beta_3(T) &=&  2 \beta_3(0) - \frac12 T  +  \bar w_3(T),
  \end{eqnarray}
while the off-diagonal components are given by stochastic integrals (involving three more independent
Wiener processes $w_{12}(T), w_{13}(T),w_{23}(T)$) of the type
\begin{equation}
  N^1_{\; \: 2}(T) - N^1_{\; \: 2}(0) \sim \int_0^T \exp ( \beta_2(T')- \beta_1(T')) dw_{12}(T') . 
  \end{equation}
  The latter stochastic
integrals converge to finite (random) limits as $T \to +\infty$.  
The existence of a finite limit follows
from the fact that the exponents entering the differential of  $  N^a_{\; \: i}(T)$ are decreasing approximately linearly in $T$;
e.g. (recalling  that Brownian
motions $w_a(T)$ behave as $ O( T^{\frac12})$ when $T \to +\infty$)
 \begin{equation}
\beta_2(T)- \beta_1(T)=  \beta_2(0)- \beta_1(0) - \frac14 T + \frac12 \bar w_2(T) - \frac12 \bar w_1(T) \approx -\frac14 T +O(T^{\frac12}) . 
 \end{equation}

We note in passing that the asymptotic (large $T$) stochastic evolution of the metric $ \bar g_{ij}(T)$
 is reminiscent of the Belinsky-Khalatnikov-Lifshitz \cite{Belinsky:1970ew}
dynamics of the spatial metric $  g_{ij}(\tau)$ (considered at a fixed spatial point) near a cosmological singularity. Indeed, when parametrizing small cosmic proper times $t \to 0$
by  $\tau = \int \frac{dt}{\sqrt{g}} \sim - \ln t \to + \infty$, and using an Iwasawa parametrization, the off-diagonal components 
$N^a_{\; \: i}(\tau) $ of the metric have finite limits as $\tau \to  + \infty$,
while  the diagonal components of the metric, 
$a_1^2= e^{2 \beta_1}$, $a_2^2= e^{2 \beta_2}$,  $a_3^2= e^{2 \beta_3}$, 
 have a stochastic behavior made of an infinite succession of Kasnerlike periods during which
 the  $ \beta_a$'s evolve approximately linearly with $\tau$,
 say $\beta_a\approx \beta_a^0 + c_a  \tau$, with velocities $c_a =O(1)$, see, e.g., \cite{Damour:2002et,Damour:2007nb}.

When going back from the heat-diffusion of the rescaled metric  $\bar g_{ij}$  over $S$ with respect to the rescaled time $T$, 
Eq.~\eqref{T}, to the stochastic evolution of the original spatial metric $g_{ij}({\bf x}, t)=  e^{2 H t} \hat g_{ij}({\bf x}, t)$
in synchronous gauge, the final result is that the spatial metric at each spatial point would stochastically wander in the
following approximate way
 \begin{eqnarray}
 g_{ij}({\bf x}, t) &\approx&  e^{2 H t} \left( e^{ \frac12 T + 2 \beta_1^0+ \bar w_1(T)}  N^1_{\; \: i}(T)  N^1_{\; \: j}(T) +  e^{2 \beta_2^0+ \bar w_2(T)}   N^2_{\; \: i}(T)  N^2_{\; \: j}(T)  \right.\nonumber\\
  &+&  \left. e^{- \frac12 T + 2 \beta_3^0 + \bar w_3(T)}  N^3_{\; \: i}(T)  N^3_{\; \: j}(T)\right)\,,
  \end{eqnarray}
  where we recall that $T \sim  \left(\frac{\kappa H}{2 \pi}   \right)^2 \, H t $.
  
  This quantum asymptotic behavior differs from the classical cosmic-no-hair asymptotic behavior of Eqs. \eqref{ds2}, \eqref{exp}:
  \begin{equation}
  g_{ij}({\bf x}, t) \approx  e^{2 H t} a_{ij}({\bf x})\,,
  \end{equation}
  in that the time-independent limiting classical rescaled (unimodular) metric $a_{ij}({\bf x})$ is replaced by a (unimodular) metric whose
  diagonal elements (in an Iwasawa parametrization) keep evolving in an exponential way. 
  This corresponds to a {\it  quantum violation of
  the cosmic-no-hair behavior} (the rescaled metric becoming more and more anisotropic in the future).
  However, we note that the anisotropic deviations from a uniformly expanding de Sitter metric are fractionally
  small. Indeed, we have an anisotropic Hubble expansion for the diagonal 
  components $a_a^2$ of the type $a_a \approx \exp H_a t$, where $a=1,2,3$ and 
  \begin{equation}
H_a = H +  c_a \left(\frac{\kappa H}{2 \pi}   \right)^2 \, H\,,
 \end{equation}
 with $c_a \propto (+\frac14, 0, -\frac14)$ (modulo an overall factor of order unity).
  Here the dimensionless (one-loop) factor $\left(\frac{\kappa H}{2 \pi}   \right)^2 $ (which is proportional to $G$)
  must be assumed to be small in the present treatment.
  Note also that, in the approximation used here, we have no indication of the presence of a
  quantum back reaction slowing down the exponential inflation (as was 
  suggested by several authors, notably by Polyakov \cite{Polyakov:1982ug}, and by Tsamis and Woodard \cite{Tsamis:1996qq}). 
  This is not unexpected as our approximation
  is a resummed version of one-loop effects, while the perturbative computations of Ref. \cite{Tsamis:1996qq}
  (see also \cite{Miao:2024shs,Woodard:2025smz}) found a slowing down of inflation at the two-loop level,
  i.e., $H_{\rm effective} \sim H \left[ 1- c  \left(\frac{\kappa H}{2 \pi}   \right)^4 (H t)^2 \right]$ with a positive
  coefficient $c= O(1)$.
 
  I leave to future work the task\footnote{I also leave to future work the task of putting on a firmer ground
  the sketchy derivation given above,  with precise numerical estimates of the various $O(1)$ factors 
  that entered it.} of comparing  the results sketched above to the recent  leading-logarithm
  quantum gravity results of Ref. \cite{Miao:2024shs}.  This task poses the delicate problem of defining
  observable quantities that could be compared in both approaches (and that are physically meaningful
  within a quantum cosmological spacetime).

 {\bf Summarizing:}  a possibly too naive, and certainly preliminary, application of Starobinsky's stochastic approach to IR
 quantum-gravity effects   in an inflationary spacetime suggests a picture where the low-frequency part
 of the metric (coarse-grained on Hubble scales) undergoes (modulo the overall
 exponential expansion factor $e^{2 H t}$) a Brownian motion in the 5-dimensional symmetric space 
$ SL(3)/SO(3)$. This leads to a quantum violation of
  the cosmic-no-hair behavior  in which the rescaled metric $ \bar g_{ij}({\bf x}, t)$ 
  does not tend to a finite limit in the asymptotic future, but becomes more and more anisotropic in
  a way reminiscent of the  Belinsky-Khalatnikov-Lifshitz behavior near a singularity.

\begin{acknowledgement}
It is a pleasure to thank Sasha Polyakov and Richard Woodard for enlightening discussions about IR effects in de Sitter spacetimes,
and Yvain Bruned and Laurence Field for informative exchanges about stochastic differential equations.
The present work was partly supported by the {\it ``2021 Balzan Prize
 for Gravitation: Physical and Astrophysical Aspects"}.
\end{acknowledgement}
%



\begin{thebibliography}{99}

\bibitem{Starobinsky:1982ee}
A.~A.~Starobinsky,
``Dynamics of Phase Transition in the New Inflationary Universe Scenario and Generation of Perturbations,''
Phys. Lett. B \textbf{117}, 175-178 (1982)

\bibitem{Vilenkin:1983xq}
A.~Vilenkin,
``The Birth of Inflationary Universes,''
Phys. Rev. D \textbf{27}, 2848 (1983)

\bibitem{Linde:1986fd}
A.~D.~Linde,
``Eternally Existing Selfreproducing Chaotic Inflationary Universe,''
Phys. Lett. B \textbf{175}, 395-400 (1986)

\bibitem{Starobinsky:1986fx}
A.~A.~Starobinsky,
``Stochastic de Sitter (inflationary) stage in the early universe,"
Lect. Notes Phys. \textbf{246}, 107-126 (1986)

\bibitem{Starobinsky:1994bd}
A.~A.~Starobinsky and J.~Yokoyama,
``Equilibrium state of a selfinteracting scalar field in the De Sitter background,''
Phys. Rev. D \textbf{50}, 6357-6368 (1994)
[arXiv:astro-ph/9407016 [astro-ph]].

\bibitem{Cruces:2022imf}
D.~Cruces,
``Review on Stochastic Approach to Inflation,''
Universe \textbf{8}, no.6, 334 (2022)
[arXiv:2203.13852 [gr-qc]].

\bibitem{Woodard:2025cez}
R.~P.~Woodard,
``Recent Developments in Stochastic Inflation,''
[arXiv:2501.15843 [gr-qc]].

\bibitem{Polyakov:2012uc}
A.~M.~Polyakov,
``Infrared instability of the de Sitter space,''
[arXiv:1209.4135 [hep-th]].

\bibitem{Anderson:2017hts}
P.~R.~Anderson, E.~Mottola and D.~H.~Sanders,
``Decay of the de Sitter Vacuum,''
Phys. Rev. D \textbf{97}, no.6, 065016 (2018)
[arXiv:1712.04522 [gr-qc]].

\bibitem{Akhmedov:2019cfd}
E.~T.~Akhmedov, U.~Moschella and F.~K.~Popov,
``Characters of different secular effects in various patches of de Sitter space,''
Phys. Rev. D \textbf{99}, no.8, 086009 (2019)
[arXiv:1901.07293 [hep-th]].

\bibitem{Akhmedov:2019esv}
E.~T.~Akhmedov, K.~V.~Bazarov, D.~V.~Diakonov, U.~Moschella, F.~K.~Popov and C.~Schubert,
``Propagators and Gaussian effective actions in various patches of de Sitter space,''
Phys. Rev. D \textbf{100}, no.10, 105011 (2019)
[arXiv:1905.09344 [hep-th]].

\bibitem{Cohen:2020php}
T.~Cohen and D.~Green,
``Soft de Sitter Effective Theory,''
JHEP \textbf{12}, 041 (2020)
[arXiv:2007.03693 [hep-th]].

\bibitem{Miao:2024shs}
S.~P.~Miao, N.~C.~Tsamis and R.~P.~Woodard,
``Leading Logarithm Quantum Gravity,''
[arXiv:2409.12003 [gr-qc]].

\bibitem{Woodard:2025smz}
R.~P.~Woodard,
``Resummations for Inflationary Quantum Gravity,''
[arXiv:2501.05077 [gr-qc]].

\bibitem{Starobinsky:1982mr}
A.~A.~Starobinsky,
``Isotropization of arbitrary cosmological expansion given an effective cosmological constant,''
JETP Lett. \textbf{37}, 66-69 (1983)

\bibitem{Starobinsky:1979ty}
A.~A.~Starobinsky,
``Spectrum of relict gravitational radiation and the early state of the universe,''
JETP Lett. \textbf{30}, 682-685 (1979)

\bibitem{Kiefer:1998qe}
C.~Kiefer, D.~Polarski and A.~A.~Starobinsky,
``Quantum to classical transition for fluctuations in the early universe,''
Int. J. Mod. Phys. D \textbf{7}, 455-462 (1998)
[arXiv:gr-qc/9802003 [gr-qc]].

\bibitem{Henneaux:1981su}
M.~Henneaux, M.~Pilati and C.~Teitelboim,
``Explicit Solution for the Zero Signature (Strong Coupling) Limit of the Propagation Amplitude in Quantum Gravity,''
Phys. Lett. B \textbf{110}, 123-128 (1982)

\bibitem{Ito1950}
K. It\^o, 
``Stochastic differential equations in a differentiable manifold",
Nagoya Math. J. \textbf{1}, 35-47 (1950)

\bibitem{Damour:2011yk}
T.~Damour and P.~Spindel,
``Quantum Einstein-Dirac Bianchi Universes,''
Phys. Rev. D \textbf{83}, 123520 (2011)
[arXiv:1103.2927 [gr-qc]].

\bibitem{Malliavin1974}
M-P. Malliavin and P. Malliavin,  ``Factorisations et lois limites de la diffusion horizontale au-dessus d'un espace riemannien sym\'etrique", Lec. Notes Math., \textbf{404}, 166-217 (1974).

\bibitem{Anker1992}
Jean-Philippe Anker and  Alberto G. Setti, 
``Asymptotic finite propagation speed for heat diffusion on certain Riemannian manifolds",
Journal of Functional Analysis, \textbf{103}, 50 (1992)

\bibitem{Matsumoto2022}
H. Matsumoto, and J.  Otani,
``Laplacian and Brownian motion on positive definite matrices, revisited",
Statistics and Probability Letters, \textbf{ 193}, 109696 (2022).

\bibitem{Belinsky:1970ew}
V.~A.~Belinsky, I.~M.~Khalatnikov and E.~M.~Lifshitz,
``Oscillatory approach to a singular point in the relativistic cosmology,''
Adv. Phys. \textbf{19}, 525-573 (1970)

\bibitem{Damour:2002et}
T.~Damour, M.~Henneaux and H.~Nicolai,
``Cosmological billiards,''
Class. Quant. Grav. \textbf{20}, R145-R200 (2003)
[arXiv:hep-th/0212256 [hep-th]].

\bibitem{Damour:2007nb}
T.~Damour and S.~de Buyl,
``Describing general cosmological singularities in Iwasawa variables,''
Phys. Rev. D \textbf{77}, 043520 (2008)
[arXiv:0710.5692 [gr-qc]].

\bibitem{Polyakov:1982ug}
A.~M.~Polyakov,
``Phase transitions and the Universe,"
Sov. Phys. Usp. \textbf{25}, 187 (1982)

\bibitem{Tsamis:1996qq}
N.~C.~Tsamis and R.~P.~Woodard,
``Quantum gravity slows inflation,''
Nucl. Phys. B \textbf{474}, 235-248 (1996)
[arXiv:hep-ph/9602315 [hep-ph]].


\end{thebibliography}
\end{document}